\documentclass[runningheads]{llncs}
\usepackage{graphicx}
\usepackage{cite} 
\usepackage{times}
\usepackage{epsfig}
\usepackage{amsmath}
\usepackage{amssymb}
\usepackage{mwe}
\usepackage{acro}
\usepackage{amssymb}
\usepackage{xcolor,colortbl}
\usepackage{tabularx}
\usepackage{relsize}
\usepackage{pifont}
\usepackage{booktabs} 
\usepackage{multirow}
\usepackage{multicol}
\usepackage{adjustbox}
\usepackage{float}

\DeclareAcronym{ROI}{
short=ROI,
long=region of interest,
}

\DeclareAcronym{IOU}{
short=IOU,
long=intersection over union,
}

\DeclareAcronym{cIOU}{
short=cIOU,
long=circle intersection over union,
}

\DeclareAcronym{DoF}{
short=DoF,
long=degrees of freedom,
}

\begin{document}

\title{EasierPath: An Open-source Tool for Human-in-the-loop Deep Learning of Renal Pathology}


\author{Zheyu Zhu\inst{1} \and 
Yuzhe Lu \inst{1} \and
Ruining Deng \inst{1} \and
Haichun Yang \inst{2} \and
Agnes B. Fogo \inst{2} \and
Yuankai Huo* \inst{1}}

% index{Zhu, Zheyu}
% index{Lu, Yuzhe}
% index{Deng, Ruining}
% index{Yang, Haichun}
% index{Fogo, Agnes}
% index{Huo, Yuankai}

%\author{submission ID: 4}
%
% \authorrunning{H. Yang et al.}

\institute{Department of Electrical Engineering and Computer Science, Vanderbilt University, Nashville, TN, USA 37235 \and
Department of Pathology, Microbiology \& Immunology, Vanderbilt University Medical Center, Nashville, TN, USA 37232} 
% }
% \email{yuankai.huo@vanderbilt.edu}}

\maketitle

\begin{abstract}
Considerable morphological phenotyping studies in nephrology have emerged in the past few years, aiming to discover hidden regularities between clinical and imaging phenotypes. Such studies have been largely enabled by deep learning based image analysis to extract sparsely located targeting objects (e.g., glomeruli) on high-resolution whole slide images (WSI). However, such methods need to be trained using labor-intensive high-quality annotations, ideally labeled by pathologists. Inspired by the recent ``human-in-the-loop'' strategy, we developed EasierPath, an open-source tool to integrate human physicians and deep learning algorithms for efficient large-scale pathological image quantification as a loop. Using EasierPath, physicians are able to (1) optimize the recall and precision of deep learning object detection outcomes adaptively, (2) seamlessly support deep learning outcomes refining using either our EasierPath or prevalent ImageScope software without changing physician's user habit, and (3) manage and phenotype each object with user-defined classes. As a user case of EasierPath, we present the procedure of curating large-scale glomeruli in an efficient human-in-the-loop fashion (with two loops). From the experiments, the EasierPath saved 57\% of the annotation efforts to curate 8,833 glomeruli during the second loop. Meanwhile, the average precision of glomerular detection was leveraged from 0.504 to 0.620. The EasierPath software has been released as open-source to enable the large-scale glomerular prototyping. The code can be found in \url{https://github.com/yuankaihuo/EasierPath}.

\end{abstract}
\section{Introduction}
In the past decades, the digital image processing algorithms have been widely applied to digital renal pathology images, especially with advanced deep learning algorithms~\cite{servais2011interstitial,grimm2003computerized,kato2015segmental,klapczynski2012computer,gadermayr2016we,ginley2017unsupervised,ginley2018computational,gadermayr2019cnn,litjens2016deep,litjens2017survey,hermsen2019deep}. However, one major challenge of employing deep learning algorithms is the requirement of massive training data with manual annotation. Moreover, the annotation on pathological images requires extensive professional knowledge and resources, which is resource-intensive and tedious for pathologists. To alleviate the human efforts, the human-in-the-loop deep learning strategy has become a promising direction~\cite{lutnick2019integrated,wang2019weakly}, whose aim is to integrate human and machine intelligence for efficient deep model deployment. 

In this paper, we propose an open-source tool EasierPath, which integrates human physicians and deep learning algorithms, for efficient large-scale object detection of renal pathology (Figure 1). Briefly, CircleNet~\cite{yang2020circlenet} was employed to perform automatic glomerular detection. Then, automatic object detection results are globally curated by adjusting the optimal detection threshold for maximizing the true positive and minimizing the false positive. Next, pathologists perform manual quality assurance (QA) and correction upon detection results.  The manual QA and correction are not only enabled by using EasierPath, but also seamlessly compatible with the most prevalent commercial software ImageScope\cite{murray2011neuropathologically}. Last, glomeruli were extracted, managed, and documented by EasierPath with customized classification.

   \begin{figure} [ht]
   \begin{center}
   \includegraphics[width=12cm]{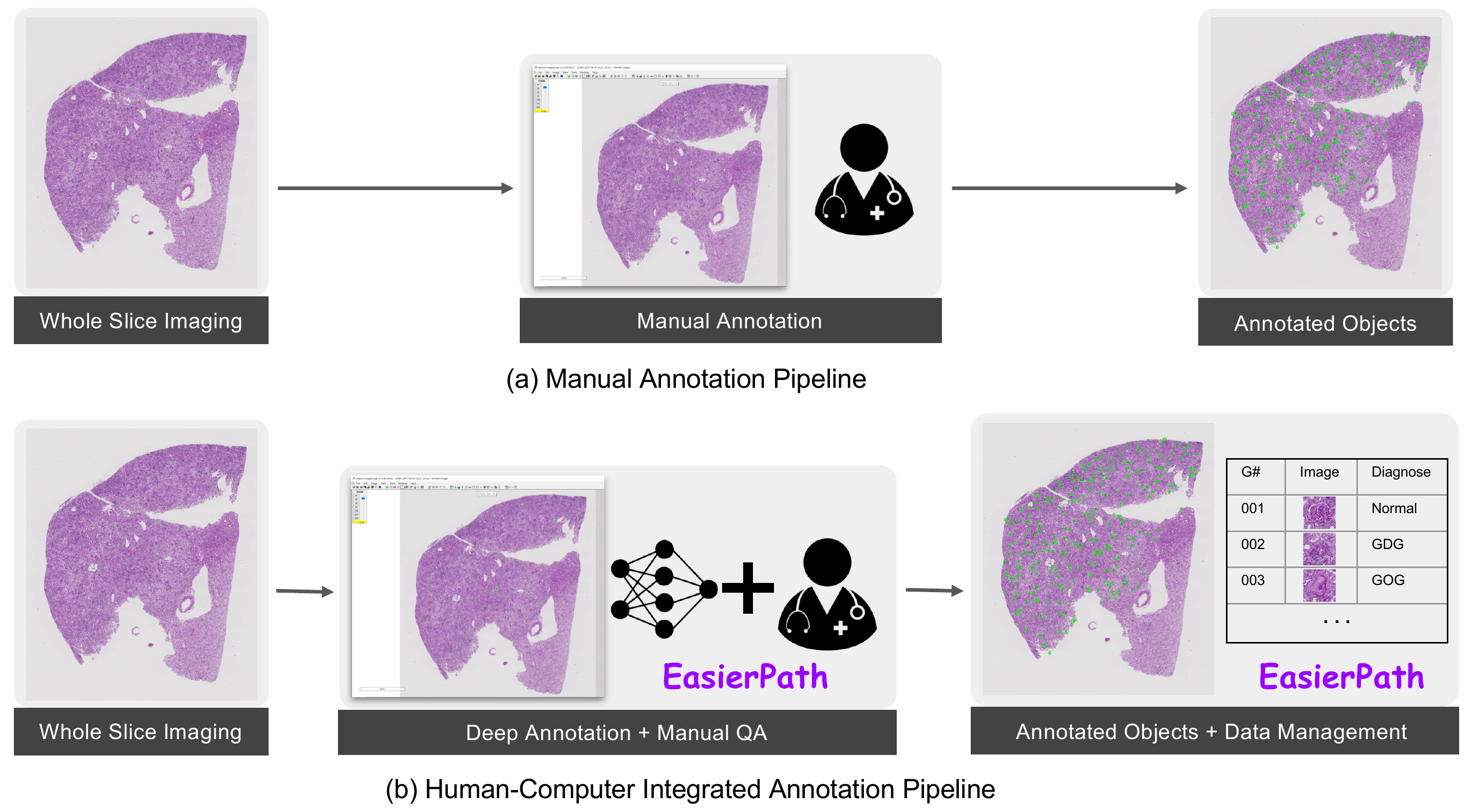}
   \end{center}
   \caption[example] 
%>>>> use \label inside caption to get Fig. number with \ref{}
   { \label{Figure_1} 
This figure illustrates two different strategies for glomerular annotation. The upper panel represents the traditional method that the entire dataset is annotated directly by human doctors. The lower panel represents the proposed human-in-the-loop framework, enabled by our EasierPath software. Such framework integrates computers and humans for more efficient data annotation and curation. (``GDG'' means Global Disappearing Glomerulosclerosis and ``GOG'' means Global Obsolescent Glomerulosclerosis)}
   \end{figure}

\begin{figure} [ht]
   \begin{center}
   \includegraphics[width=12cm]{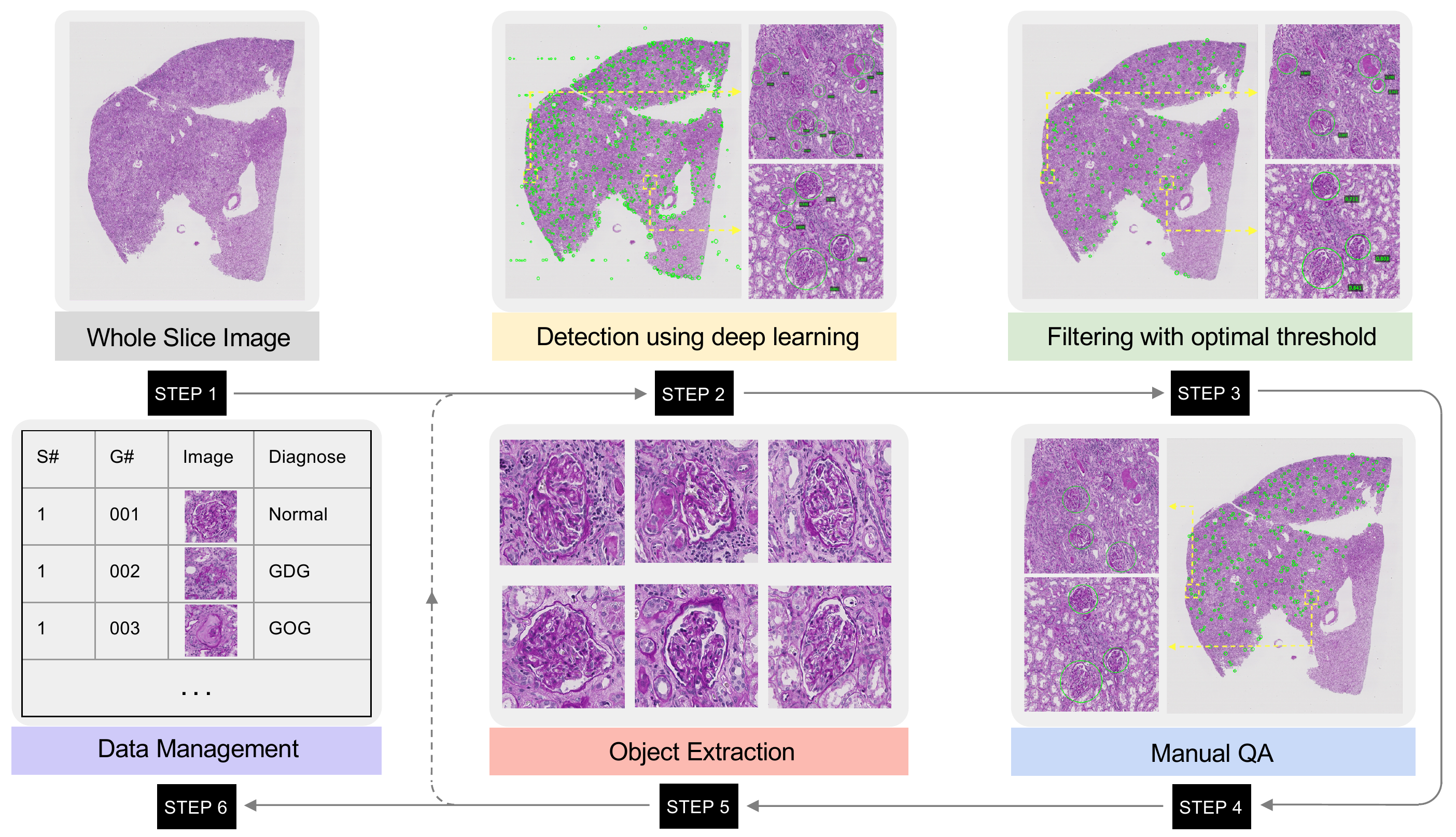}
   \end{center}
   \caption[example] 
%>>>> use \label inside caption to get Fig. number with \ref{}
   { \label{Figure_2} 
This figure illustrates the workflow of the human-in-the-loop annotation pipeline. Step 1 shows an example of an input WSI. Step 2 is a deep learning based object detection. Step 3 is the global filtering of detection results. Step 4 is the local manual QA. Step 5 and 6 show the object extraction as well as the corresponding class annotation results. The dashed line indicates that the curated data could be used to retrain the deep learning algorithm as a ``loop''.}
   \end{figure}

\section{Methods}

Figure 2 shows the entire workflow, which consists of deep learning based detection, filtering, manual QA, and data management.

\subsection{Detection using deep learning}
The input image of the entire flow is a high-resolution whole slide image (WSI). Then, the automatic glomerular detection results are achieved from CircleNet~\cite{yang2020circlenet}. The detection outcomes are saved in one XML file that contains circle location, type, and the detection score for each detected object. The detection score is a score within 0 to 1, where a larger score indicates the stronger confidence to believe the detected object is a glomerulus. The XML format file can be loaded with the corresponding intensity image in EasierPath, for further QA and data curation.

\begin{figure} [ht]
   \begin{center}
   \includegraphics[width=12cm]{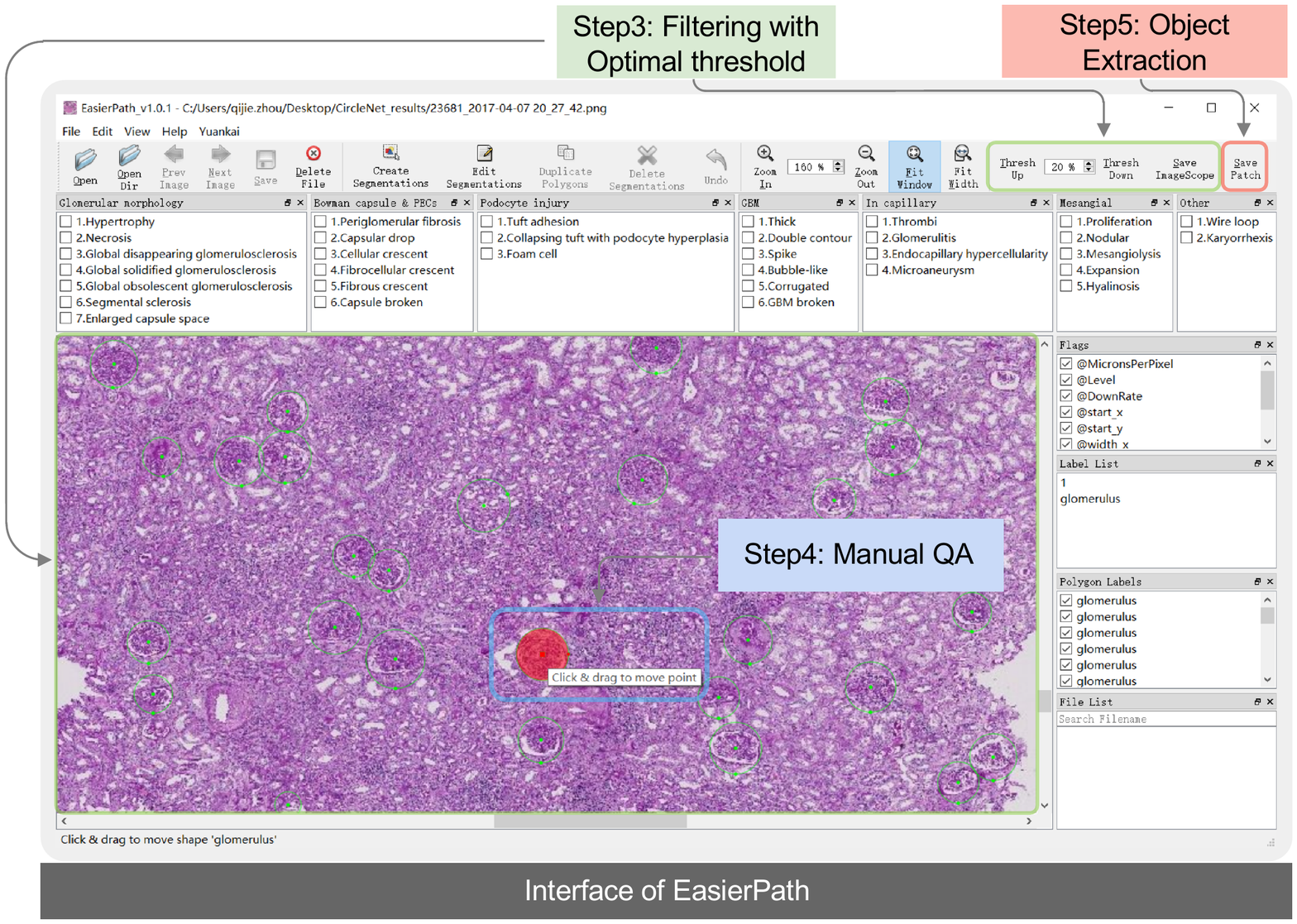}
   \end{center}
   \caption[example] 
%>>>> use \label inside caption to get Fig. number with \ref{}
   { \label{Figure_3} 
This figure presents the interface of EasierPath. Toolbox and where each step is operated. Step 3 provides buttons to control the global false positive and false negative by filtering the detection score with different thresholds. Then, the local image QA and curation (step 4) can be performed using our annotation tool. The click on ``Save Patch'' in step 5 will automatically extract all the detected objects and save each object as an individual image.}
   \end{figure}

\subsection{Filtering with optimal threshold}
After deep learning based detection, we can visualize the results by importing the XML file into the EasierPath software. Using EasierPath, we can alter the threshold of detection scores to decide which glomeruli should be kept. The threshold is adjusted by clicking the button ``Thresh Up'' or ``Thresh Down'' (Figure 3). All the circles with detection scores lower than that threshold will disappear for each threshold, leaving behind circles with detection scores higher than or equal to that threshold. The threshold adjustment allows the global balance of false positive and false negative to minimize the manual efforts for the following QA.

\subsection{Manual quality assurance}
After selecting an optimal threshold, doctors can perform manual QA using EasierPath software. As the most labor intensive step in the pipeline, to leverage the annotation efficiency for clinicians who would like to use ImageScope (\url{www.leicabiosystems.com}) without changing their user habits, we provide the alternative option to export the detection results after thresholding as an ImageScope compatible format(by simply clicking ``Save ImageScope'' button). It enables the seamless annotation format conversion between EasierPath and ImageScope. Using EasierPath or ImageScope, clinicians are able delete false positive, annotate false negative, and correct detection results.

\subsection{Improve the deep learning based detection as a loop}
Once we get the manually curated dataset, we can use those datasets as additional training data to leverage the performance of the deep learning algorithm as a ``loop'', which is the crucial idea in the human-in-the-loop design. With more training data, the performance of the deep learning network will typically be improved for the next loop. That will further help the following manual QA, upon the more accurate automatic results. Note that the Step 2 and Step 3 will not be performed during the first loop, since we don't have any annotated data to train the detection model.

\begin{figure} [ht]
   \begin{center}
   \includegraphics[width=12cm]{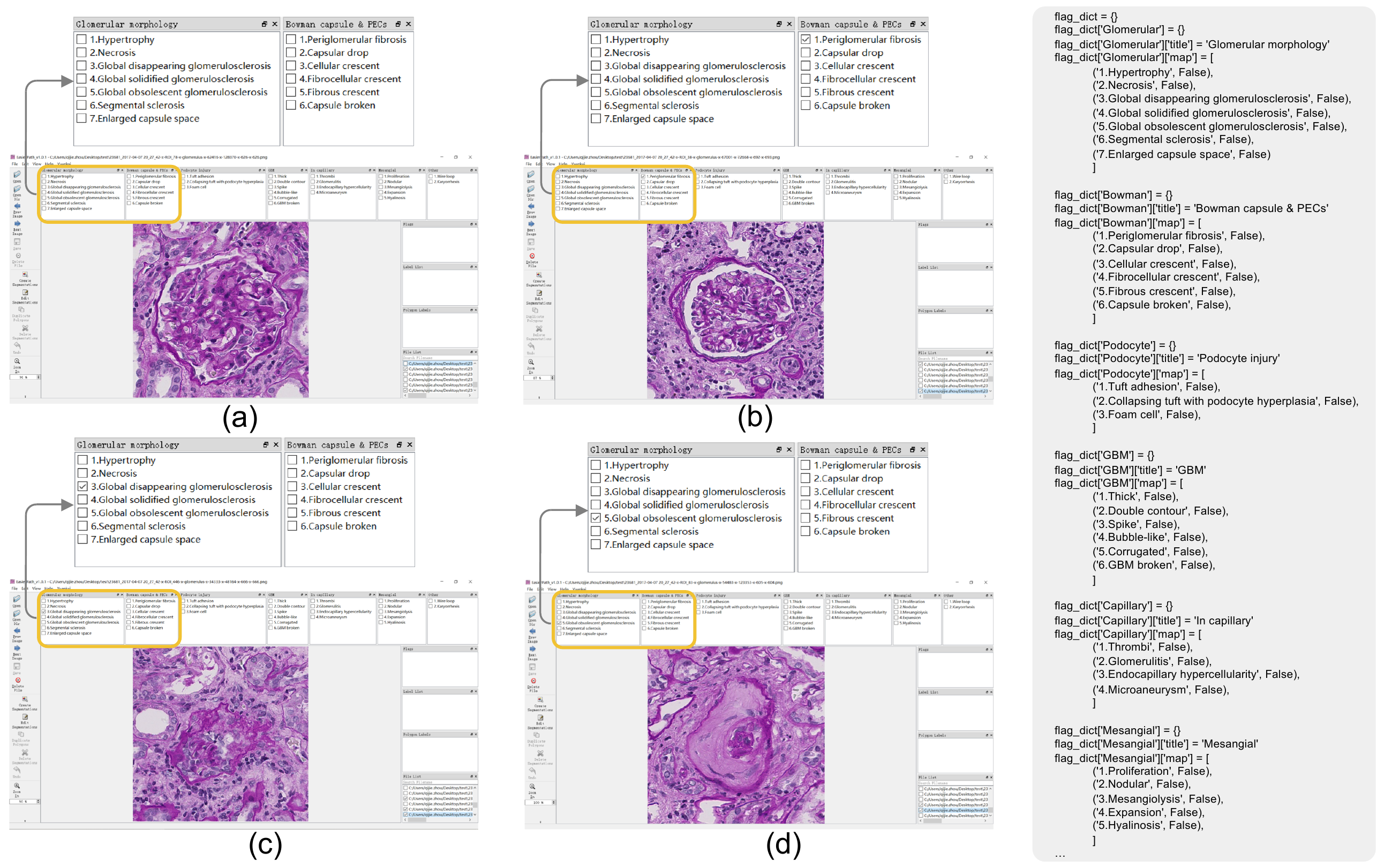}
   \end{center}
   \caption[example] 
   {\label{Figure_4} 
    This figure shows the interface of EasierPath for manual class annotation. (a)-(d) show examples of how to use EasierPath to perform glomerular classification. The definition of the classes can be easily changed in the configuration file (right panel) by clinicians without programming skills.}
%>>>> use \label inside caption to get Fig. number with \ref{}
\end{figure}

\subsection{Object extraction and management}
After completing the manual QA, all targeting objects have been annotated. As the histology images are typically high resolution Gigabytes images, while the objects of interest typically only exhibit small portions of the entire image. Therefore, we could only extract, manage, and save those meaningful pixels, to accelerate the following secondary analysis, data retrieval, and model training.

First, all the image patches that contain the targeting objects can be saved as individual images by a single click (Figure 4). Then, the cropped patch samples can be loaded and labeled efficiently using the same EasierPath software. The physicians are able to define the categories of labels by editing a configuration file conveniently without programming skills. Once the configuration is confirmed, all the image patches can be loaded to the EasierPath and be annotated efficiently (Figure 4). After clicking, the annotations will be saved into a json format file. The json files and the patch files will be saved as a database for future utilization. For instance, we can efficiently extract all glomeruli from the database with ``global glomerulosclerosis'' in the future, we would like to investigate such phenotype or train a machine learning based classification method.

\section{Data}

WSI from renal biopsies and human kidney nephrectomy tissues were utilized for performing the glomerular quantification at the first and second loop respectively. In the first loop, the renal biopsies were quantified, whose kidney biopsy tissues were routinely processed, paraffin-embedded, and 2 $\mu$m thickness sections cut and stained with hematoxylin and eosin (HE), periodic acid–Schiff (PAS) or Jones. In the second loop, the human nephrectomy tissues were quantified, whose tissues were routinely processed, paraffin-embedded, and 3 $\mu$m thickness sections cut and stained with PAS. The data were de-identified, and studies were approved by the Institutional Review Board (IRB). After the first loop, the CircleNet was trained by 704 glomeruli from 42 biopsy samples to perform initial detection for the second loop. In the second loop, 7,449 glomeruli from 18 human nephrectomy images were curated by using both automatic detection and manual QA in EasierPath framework. After the second loop, all curated data are used as training and validation data for retraining the CircleNet. We manually annotate 1384 glomeruli from five untouched human nephrectomy images as testing data to evaluate the performance of detection after the first and second loops.

\begin{table}
    \centering
    \caption{Detection results with CircleNet}
    \begin{tabular}{c c c c c c c c}
         \hline
          & Loop & $AP$ & $AP_{50}$ & $AP_{75}$ & $AP_{S}$ & $AP_{M}$ & $AP_{L}$\\
         \hline
         CircleNet \quad & Loop=1 \quad & 0.504 & 0.729 & 0.511 & 0.363 & 0.721 & 0.625\\
         CircleNet \quad & Loop=2 \quad & 0.620 & 0.915 & 0.602 & 0.531 & 0.756 & 0.668\\
         \hline
    \end{tabular}
    \label{CircleNetOriginalTable}
\end{table}

\section{Experiments and Results}
\subsection{Labor cost analysis}
In the second loop, we randomly chose one complete human nephrectomy image to evaluate the labor cost between the two strategies in Figure 1. Using pure manual annotation, it took about 7 seconds per glomeruli by a renal pathologist with more than 20 years of experience. It took about 3 seconds per glomeruli for the same pathologists using the proposed EasierPath pipeline with the initial CircleNet (after the first loop). The temporal gap between the pure manual annotation and using EasierPath pipeline was more than two weeks to avoid the annotator remember the same human nephrectomy image. From the test, $57.1\%$ of the manual efforts are reduced using the proposed framework.

\begin{figure}
   \begin{center}
   \includegraphics[width=10cm]{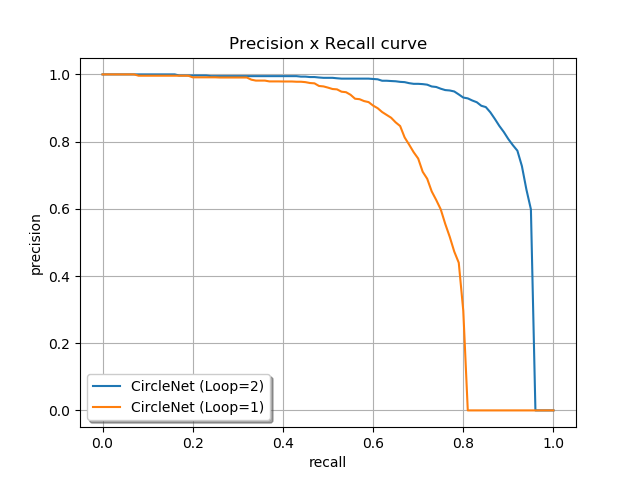}
   \end{center}
   \caption[example] 
   {\label{Figure_5} 
    This figure shows the precision-recall curves from the CircleNet detection after the first and second loop respectively.}
%>>>> use \label inside caption to get Fig. number with \ref{}
\end{figure}

\subsection{Detection performance}

1384 glomeruli from five untouched human nephrectomy images were manually annotated by another independent annotator as testing data to evaluate the performance of deep learning detection after the first and second loops. We report average precision (AP) related canonical detection metrics overall Intersection over Union (IOU) thresholds (Table 1), which shows the result of CircleNet after the first loop (Loop=1) and the second loop (Loop=2). 

According to Table 1, CircleNet after the second loop achieved a better accuracy for all types of APs. Especially when area = small, average precision for CircleNet after the second loop was approximately 46.28\% higher than the performance after the first loop. When IOU = 0.5, the average precision for CircleNet achieved approximately 25.51\% improvements. Better accuracy indicated that the performance of CircleNet was leveraged after when performing more loops. 

To visualize the performance, precision-recall curves for CircleNet after the first and second loops were shown in Figure 5. Precision, the ratio of TP/(TP + FP)\cite{ozenne2015precision}, was higher for CircleNet with manual annotation for any recall value, which indicated the performance of CircleNet was leveraged from more training data with more loops. As recall value represented the ratio of FN/(TN + FN), a higher recall curve was obtained after the second loop. When the precision rate was about 0.8, the recall rate for CircleNet (Loop = 1) was approximately 0.68, while the recall rate for CircleNet (Loop = 2) was about 0.9, which equaled to about 32.35\% improvements. Using human in the loop for only two loops, CircleNet with manual annotation has provided decent performance on detection of glomeruli.

\section{Conclusion}
In this paper, we introduced EasierPath, an efficient large-scale pathological image quantification tool with human-in-the-loop deep learning of renal pathology. The proposed method reduced the $57\%$ labor cost for curating large-scale target objects in high-resolution pathological WSI. Meanwhile, the performance of deep learning detection was leveraged after performing each loop. Last, the EasierPath tool provided a easy-to-adapt function to curate, extract, and manage each detected object for future usage. 

\section{ACKNOWLEDGMENTS}
This work was supported by NIH NIDDK DK56942(ABF).

\bibliographystyle{splncs04}
\bibliography{paper04}

\begin{thebibliography}{10}
\providecommand{\url}[1]{\texttt{#1}}
\providecommand{\urlprefix}{URL }
\providecommand{\doi}[1]{https://doi.org/#1}

\bibitem{gadermayr2019cnn}
Gadermayr, M., Dombrowski, A.K., Klinkhammer, B.M., Boor, P., Merhof, D.: Cnn
  cascades for segmenting sparse objects in gigapixel whole slide images.
  Computerized Medical Imaging and Graphics  \textbf{71},  40--48 (2019)

\bibitem{gadermayr2016we}
Gadermayr, M., Klinkhammer, B.M., Boor, P., Merhof, D.: Do we need large
  annotated training data for detection applications in biomedical imaging? a
  case study in renal glomeruli detection. In: International Workshop on
  Machine Learning in Medical Imaging. pp. 18--26. Springer (2016)

\bibitem{ginley2017unsupervised}
Ginley, B., Tomaszewski, J.E., Yacoub, R., Chen, F., Sarder, P.: Unsupervised
  labeling of glomerular boundaries using gabor filters and statistical testing
  in renal histology. Journal of Medical Imaging  \textbf{4}(2),  021102 (2017)

\bibitem{ginley2018computational}
Ginley, B.G., Tomaszewski, J.E., Jen, K.Y., Fogo, A., Jain, S., Sarder, P.:
  Computational analysis of the structural progression of human glomeruli in
  diabetic nephropathy. In: Medical Imaging 2018: Digital Pathology. vol.
  10581, p. 105810A. International Society for Optics and Photonics (2018)

\bibitem{grimm2003computerized}
Grimm, P.C., Nickerson, P., Gough, J., McKenna, R., Stern, E., Jeffery, J.,
  Rush, D.N.: Computerized image analysis of sirius red--stained renal
  allograft biopsies as a surrogate marker to predict long-term allograft
  function. Journal of the American Society of Nephrology  \textbf{14}(6),
  1662--1668 (2003)

\bibitem{hermsen2019deep}
Hermsen, M., de~Bel, T., Den~Boer, M., Steenbergen, E.J., Kers, J., Florquin,
  S., Roelofs, J.J., Stegall, M.D., Alexander, M.P., Smith, B.H., et~al.: Deep
  learning--based histopathologic assessment of kidney tissue. Journal of the
  American Society of Nephrology  \textbf{30}(10),  1968--1979 (2019)

\bibitem{kato2015segmental}
Kato, T., Relator, R., Ngouv, H., Hirohashi, Y., Takaki, O., Kakimoto, T.,
  Okada, K.: Segmental hog: new descriptor for glomerulus detection in kidney
  microscopy image. Bmc Bioinformatics  \textbf{16}(1), ~316 (2015)

\bibitem{klapczynski2012computer}
Klapczynski, M., Gagne, G.D., Morgan, S.J., Larson, K.J., LeRoy, B.E., Blomme,
  E.A., Cox, B.F., Shek, E.W.: Computer-assisted imaging algorithms facilitate
  histomorphometric quantification of kidney damage in rodent renal failure
  models. Journal of pathology informatics  \textbf{3} (2012)

\bibitem{litjens2017survey}
Litjens, G., Kooi, T., Bejnordi, B.E., Setio, A.A.A., Ciompi, F., Ghafoorian,
  M., Van Der~Laak, J.A., Van~Ginneken, B., S{\'a}nchez, C.I.: A survey on deep
  learning in medical image analysis. Medical image analysis  \textbf{42},
  60--88 (2017)

\bibitem{litjens2016deep}
Litjens, G., S{\'a}nchez, C.I., Timofeeva, N., Hermsen, M., Nagtegaal, I.,
  Kovacs, I., Hulsbergen-Van De~Kaa, C., Bult, P., Van~Ginneken, B., Van
  Der~Laak, J.: Deep learning as a tool for increased accuracy and efficiency
  of histopathological diagnosis. Scientific reports  \textbf{6},  26286 (2016)

\bibitem{lutnick2019integrated}
Lutnick, B., Ginley, B., Govind, D., McGarry, S.D., LaViolette, P.S., Yacoub,
  R., Jain, S., Tomaszewski, J.E., Jen, K.Y., Sarder, P.: An integrated
  iterative annotation technique for easing neural network training in medical
  image analysis. Nature machine intelligence  \textbf{1}(2),  112--119 (2019)

\bibitem{murray2011neuropathologically}
Murray, M.E., Graff-Radford, N.R., Ross, O.A., Petersen, R.C., Duara, R.,
  Dickson, D.W.: Neuropathologically defined subtypes of alzheimer's disease
  with distinct clinical characteristics: a retrospective study. The Lancet
  Neurology  \textbf{10}(9),  785--796 (2011)

\bibitem{ozenne2015precision}
Ozenne, B., Subtil, F., Maucort-Boulch, D.: The precision--recall curve
  overcame the optimism of the receiver operating characteristic curve in rare
  diseases. Journal of clinical epidemiology  \textbf{68}(8),  855--859 (2015)

\bibitem{servais2011interstitial}
Servais, A., Meas-Yedid, V., No{\"e}l, L., Martinez, F., Panterne, C., Kreis,
  H., Zuber, J., Timsit, M., Legendre, C., Olivo-Marin, J., et~al.:
  Interstitial fibrosis evolution on early sequential screening renal allograft
  biopsies using quantitative image analysis. American Journal of
  Transplantation  \textbf{11}(7),  1456--1463 (2011)

\bibitem{wang2019weakly}
Wang, Y., Lu, L., Cheng, C.T., Jin, D., Harrison, A.P., Xiao, J., Liao, C.H.,
  Miao, S.: Weakly supervised universal fracture detection in pelvic x-rays.
  In: International Conference on Medical Image Computing and Computer-Assisted
  Intervention. pp. 459--467. Springer (2019)

\bibitem{yang2020circlenet}
Yang, H., Deng, R., Lu, Y., Zhu, Z., Chen, Y., Roland, J.T., Lu, L., Landman,
  B.A., Fogo, A.B., Huo, Y.: Circlenet: Anchor-free detection with circle
  representation. arXiv preprint arXiv:2006.02474  (2020)

\end{thebibliography}

\end{document}